\documentclass[prd,twocolumn,superscriptaddress]{revtex4-1}

\usepackage{graphicx, amsfonts, amsmath, amssymb, amscd,  theorem, exscale,
epic, eepic, epsfig}

\usepackage{hyperref} %

\usepackage{makeidx}
\makeindex

\newcounter{Figure}
\theoremstyle{plain}
\theoremheaderfont{\scshape}

\newtheorem{Def}{\bf Definition}

\theorembodyfont{\upshape}

\theoremheaderfont{\scshape\small} \theorembodyfont{\scshape\small}

\newcommand{\slap}{\mbox{$ \triangle \mkern -13mu / \ $}}
\newcommand{\nlap}{\mbox{$ \nabla \mkern -13mu / \ $}}
\newcommand{\dlap}{\mbox{$ div \mkern -13mu / \ $}}

\newcommand{\clap}{\mbox{$ curl \mkern -23mu / \ $}}

\newcommand{\be}{\begin{equation}}
\newcommand{\ee}{\end{equation}}
\newcommand{\bea}{\begin{eqnarray}}
\newcommand{\eea}{\end{eqnarray}}
\newcommand{\beas}{\begin{eqnarray*}}
\newcommand{\eeas}{\end{eqnarray*}}

\newcommand{\Lie}{ {\mathcal L} }

\begin{document}

\title{Answering the Parity Question for Gravitational Wave Memory}

\author{Lydia Bieri}
\email{lbieri@umich.edu}
\affiliation{Dept. of Mathematics, University of Michigan, Ann Arbor, MI 48109-1120, USA}

%%%%%%%%%%%%%%%%%%%%%%%%%%%%%%%%%%%%%%%%%%%%%%%%%%%

\date{\today}

%%%%%%%%%%%%%%%%%%%%%%%%%%%%%%%%%%%%%%%%%%%%%%%%%%%
\begin{abstract}
Memory of gravitational waves in asymptotically-flat spacetimes that are solutions of the Einstein vacuum equations 
is of purely electric parity, no magnetic parity memory can occur. 
We show this by investigating what happens for various classes of asymptotically-flat Einstein-vacuum-spacetimes. 
This is understood within such spacetimes that have been shown to be stable in the fully nonlinear regime under perturbations away from Minkwoski space. 
We also lay open the structures at the transition from spacetimes that have a well-defined memory to those for which the memory formula cannot be retrieved due to their specific asymptotic behavior and the divergence of crucial integrals. 
Moreover, for the Einstein equations coupled to other fields with a common stress-energy as well as in the cosmological setting, we find that gravitational memory is of electric parity only.

\end{abstract}

%%%%%%%%%%%%%%%%%%%%%%%%%%%%%%%%%%%%%%%%%%%%%%%%%%%%%%%%%%%%%%%%%%%%%%%%%%%%%%

\maketitle

%%%%%%%%%%%%%%%%%%%%%%%%%%%%%%%%%%%%
%%%%%%%%%%%%%%%%%%%%%%%%%%%%%%%%%%%%
\section{Introduction}
%\indent

Gravitational waves in General Relativity (GR) are known to exhibit a memory effect, that is a permanent change of the spacetime, as found by Ya. Zel'dovich and B. Polnarev in a linearized situation \cite{zeldovich} and by D. Christodoulou in the fully nonlinear setting \cite{chrmemory}. There are two types of this memory \cite{lbdg3}. 
In this article, we show that for spacetimes that are asymptotically-flat solutions of the Einstein vacuum  equations, for which stability has been proven, memory is of electric parity only, magnetic memory does not occur. 

A great breakthrough in science happened in 2015 with the first detection of gravitational waves from a binary black hole merger in the two US-based LIGO facilities \cite{ligodetect1} roughly 100 years after A. Einstein had formulated the theory of general relativity. In the meantime, the LIGO and Virgo detectors have observed further events, of which the measurement in 2017 of waves generated in a neutron star binary merger was another historic milestone \cite{ligodetect2, ligodetect3}. 

In detectors like LIGO gravitational memory will show as a permanent displacement of test masses, and in detectors like NANOGrav as a frequency change of pulsars' pulses. 

In \cite{Lasky1} P. Lasky, E. Thrane, Y. Levin, J. Blackman and Y. Chen suggest a way to detect gravitational 
wave memory with LIGO. 

Such an effect was known in a linear theory since 1974, when Ya. Zel'dovich and B. Polnarev \cite{zeldovich} 
found it. The effect was believed to be tiny. Then in 1991 D. Christodoulou \cite{chrmemory} 
derived within the full nonlinear theory such a memory that was much larger than expected. 
More work by other authors followed at the time \cite{damourbl, braginsky, braginskyg, will, thorne, thorne2, jorg}, and 
recently there has been more work by a growing number of authors 
\cite{1lpst1, 1lpst2, lbdg1,lbdg3,tolwal1,winicour, Lasky1, strominger,flanagan,favata,BGYmemcosmo1, bgsty1,twcosmo}. See the previous works for more detailed references.

It was believed that the ``linear" (= regular) effect by Zel'dovich and Polnarev and the ``nonlinear" (= null) effect by Christodoulou are two aspects of the same phenomenon. The present author with D. Garfinkle showed in \cite{lbdg3} that these are indeed two different effects, the former related to fields that do not go out to null infinity, the latter related to fields that do go out to null infinity. Garfinkle and the present author derived \cite{lbdgelmagn1} the analogs of both of these phenomena outside GR for the first time, namely for the pure Maxwell equations, which are of course linear. 
A growing community has explored other theories for analogs of the memory effect, see for instance \cite{strominger}. 

Several matter- or energy-fields coupled to the Einstein equations contribute to the null memory. In collaboration with P.Chen and S.-T. Yau we showed that for the Einstein-Maxwell system a specific component of the electromagnetic field enlarges the null memory \cite{1lpst1}, \cite{1lpst2}. With D. Garfinkle we proved that there is a contribution to the null memory from neutrino radiation \cite{lbdg1} as it occurs in a core-collapse supernova or a binary neutron star merger. For the latter, we model the neutrinos by introducing the energy-momentum tensor of a null fluid into the Einstein equations. 

As there is a large literature on various aspects of memory or its analogs in other theories, we refer to the above-mentioned articles for detailed references. We do not discuss lower order effects such as angular momentum memory. This and many other aspects of gravitational waves are most interesting. In this article, we concentrate on the two types of gravitational wave memory that dominate the lower order effects. 

Memory, as known in the pioneering articles, is of electric parity. 
Recently, there has been a lot of discussion whether memory can also exhibit magnetic parity. In the following, we show that for spacetimes that are asymptotically-flat (AF) solutions of the Einstein vacuum (EV) equations, for which stability has been proven, memory is of electric parity only, thus magnetic memory does not occur. Moreover, this is still true for most coupled Einstein-matter systems. However, an ``unusual" form of stress-energy could in principle contribute to magnetic ordinary memory. An interesting example was produced by R. Wald and G. Satishchandran. 
%At this point, no evolution equations are known for such matter. 

Isolated gravitating systems in GR are described by asymptotically-flat solutions of the Einstein vacuum equations, or Einstein equations coupled to corresponding matter or energy. This has been understood in detail (in the fully nonlinear regime) through the proofs of global nonlinear stability of Minkowski space. Here, small AF initial data (controlled via weighted Sobolev norms) evolve under the EV equations to globally AF spacetimes that are causally geodescially complete (thus without any singularities). The first such global proof was given by D. Christodoulou and S. Klainerman in \cite{sta}. It was generalized to Einstein-Maxwell equations by N. Zipser in \cite{zip}, \cite{zip2} and by the present author in \cite{lydia1}, \cite{lydia2} to the borderline case for the EV equations assuming one less derivative and less fall-off 
by one power of $r$ than in \cite{sta} obtaining the borderline case in view of decay in power of $r$, indicating that 
the conditions in our main theorem on the decay at infinity of the initial data are sharp. Many proofs or partial results have been obtained by many authors in various directions. The above results are obtained via geometric-analytic invariant methods and solve the equations in full nonlinearity. (That is, no approximation was used, all the results are complete.) These proofs not only establish stability for large classes of important physical systems, but they also provide an exact knowledge of these spacetimes, including the null asymptotic behavior and gravitational radiation. Whereas the smallness of the initial data was required to establish existence and uniqueness of solutions, the null asymptotic behavior is largely independent from the smallness. Thus, we can allow for large data like black holes and derive results for these at null infinity, including radiation and memory. 
One may want to take a double-null foliation near scri and find that the asymptotic results still hold for the portion of null infinity that remains in these spacetimes with large data.

We look at the most important classes of AF spacetimes. 
We refer to the situation as in \cite{sta} by D. Christodoulou and S. Klainerman as (CK), and 
the most general situation proven to be stable in \cite{lydia1}, \cite{lydia2} by the present author is denoted as (B). 
As the data in (B) decay very slowly to Minkowski at infinity, 
the description of the {asymptotic behavior} of the 
curvature components is less precise than  
in (CK). 
However, it is as precise as it can be for these spacetimes. 
Moreover, we use (B2) to refer to the case treated in \cite{lydia3} by the present author. 
In (B2) we have slightly stronger decay on the data than in (B), but are still more general than (CK). Stability of (B2) is implied by (B), but more can be retrieved from null infinity due to stronger decay of the data. In (CK) as well as in the more general (B2) ordinary and null memory can be computed and are of electric type, magnetic memory does not occur. In (B) some quantities have non-trivial limits at null infinity and corresponding structures are obtained, whereas others do not have any limits. Due to the very general setting, a crucial integral in the memory formula diverges. Thus, no finite memory in the case (B). More can be said: even though null infinity is too rough to extract this information on memory, a crucial local relation shows that the magnetic part of the curvature that would be related to magnetic memory has to decay in the optical function $u$. 
The weighted Sobolev norms to control the data in all these cases are very different. 

The proofs of stability, \cite{sta}, \cite{lydia1}, \cite{lydia2}, \cite{lydia3}, are mathematically involved and rely on the investigation of analytic and geometric structures of solutions of the Einstein equations. In this paper, we build on these results to show that gravitational memory is of electric type only. 
For the purpose of this article we do not need the methods that led to these results but make use of their physical relevance. Here, we give a self-contained description of these spacetimes from which we derive our new results in a straightforward way.

\section{Equations and Spacetime Structures}

We consider the Einstein vacuum equations 
\be \label{Ricvac0}
R_{\mu \nu} \ = \ 0  \ , 
\ee 
for $\mu, \nu = 0,1,2,3$.

$(M, g)$ denotes our solution spacetimes. 
We denote by $t$ a maximal time function and by $u$ the optical function, $H_t$ will be the spacelike slices given by the $t$-foliation and $C_u$ the outgoing null hypersurfaces of the $u$-foliation. We denote the corresponding intersection by $S_{t,u} = H_t \cap C_u$. Quantities with a bar refer to $H_t$. For now, the zero-coordinate will be the time-coordinate, and indices $1, 2, 3$ refer to spatial coordinates. 

The Weyl tensor $W_{\alpha \beta \gamma \delta}$ is decomposed into its electric and magnetic parts, which are defined by
\begin{eqnarray}
{E_{ab}} := {W_{atbt}} \label{Wel1}
\\
{H_{ab}} := {\textstyle {1 \over 2}} {{\epsilon ^{ef}}_a}{W_{efbt}} \label{Wma1}
\end{eqnarray}
Here $\epsilon _{abc}$ is the spatial volume element and is related to the spacetime volume element by
${\epsilon_{abc}} = {\epsilon_{tabc}}$. The electric part of the Weyl tensor is the crucial ingredient in the 
equation governing the distance between two objects in free fall. In particular, their spatial separation $\Delta {x^a}$ is 
\begin{equation}
{\frac {{d^2}\Delta {x^a}} {d{t^2}}} = - {{E^a}_b}\Delta {x^b}
\label{geodev}
\end{equation}

In order to solve the initial-value problem for the Einstein vacuum equations, we have to specify initial data. 
We are going to consider three large classes of initial data yielding corresponding classes of spacetimes.  

\begin{Def} (Christodoulou-Klainerman (CK), \cite{sta}) \label{intSAFCK} 
(SAFCK) 
We define a strongly asymptotically flat initial data set in the sense 
of \cite{sta} and in the following denoted 
by SAFCK initial data set, to be 
an initial data set $(H, \bar{g}, k)$, where 
$\bar{g}$ and $k$ are sufficiently smooth and there exists a coordinate system 
$(x^1, x^2, x^3)$ defined in a neighborhood of infinity such that, \\
as 
$r = (\sum_{i=1}^3 (x^i)^2 )^{\frac{1}{2}} \to \infty$, 
$\bar{g}_{ij}$ and $k_{ij}$ are: 
\bea
\bar{g}_{ij} \ & = & \ (1 \ + \ \frac{2M}{r}) \ \delta_{ij} \ + \ o_4 \ (r^{- \frac{3}{2}}) \label{safg} \\ 
k_{ij} \ & = & \  o_3 \ (r^{- \frac{5}{2}}) \ ,  \label{safk}
\eea
where $M$ denotes the mass. 
\end{Def}
In their work, weighted Sobolev norms of appropriate energies are assumed to be controlled. 
This induces the above class of initial data.

\begin{Def} (Bieri (B), \cite{lydia1}, \cite{lydia2}) \label{intAFB} 
(AFB) 
We define an asymptotically flat initial data set to be a 
AFB initial data set, if it is 
an asymptotically flat initial data set $(H_0, \bar{g}, k)$, 
where $\bar{g}$ and $k$ are sufficiently smooth 
and 
for which there exists a coordinate system $(x^1, x^2, x^3)$ in a neighborhood of infinity such that with 
$r = (\sum_{i=1}^{3} (x^i)^2 )^{\frac{1}{2}} \to \infty$, it is:  
\bea
\bar{g}_{ij} \ & = & \ \delta_{ij} \ + \ 
o_3 \ (r^{- \frac{1}{2}}) \label{afgeng}  \\
k_{ij} \ & = & \ o_2 \ (r^{- \frac{3}{2}})    \ .    \label{afgenk}  
\eea
\end{Def} 
Here, other weighted Sobolev norms of appropriate energies are assumed to be controlled, 
yielding this class of initial data.

As a consequence from imposing less conditions on the data in (B), the 
{ spacetime curvature is not in $L^{\infty}(M)$}, as opposed to (CK). 
(B) only controls one derivative of the curvature (Ricci) in $L^2 (H)$.

Moreover, in \cite{lydia1}, \cite{lydia2}, {energy} and {linear momentum} are {well-defined and conserved}, 
whereas the {(ADM) angular momentum} is {not defined}. 
This is different to the situation investigated in \cite{sta}, where 
all these quantities are well-defined and conserved.

We work with the null frame $e_4, e_3, e_2, e_1$. That is, $e_4$ and $e_3$ form a 
null pair which is supplemented by $e_A, \ A = 1, 2$, a local 
frame field for $S_{t,u} = H_t \cap C_u $. Given this null pair, $e_3$ and $e_4$, we
can define the tensor 
of projection from the tangent space of $M$ to that of $S_{t,u}$.
\[  \Pi^{\mu \nu} = g^{\mu \nu} + 
\frac{1}{2}(e_4^{\nu}e_3^{\mu}+e_3^{\nu}e_4^{\mu}). \] 
Denote by $N$ the outward unit normal vector of $S_{t,u}$ in $H_t$ and by $T$ the future-directed unit normal to $H_t$. 
Then we see that $e_3 = T - N$ is an incoming null vectorfield, and $e_4 = T + N$ an outgoing null vectorfield. 
We will make use of $N = a^{-1} \frac{\partial}{\partial u}$ with lapse $a = |\nabla u|^{-1}$. 
To denote operators on the surfaces $S_{t,u}$ we use a slash, thus $\dlap$, $\clap \ $ are the corresponding divergence and curl operators, respectively. 
For a $p$-covariant tensor field $t$ that is tangent to $S$ we denote by $D_4 t$ and $D_3 t$ the projections to $S$ of the Lie derivatives $\Lie_{e_4} t$, respectively $\Lie_{e_3}t$. 

%We define
%\[  \theta_{AB} =  \langle \nabla_{A} N, e_B \rangle. \]

%The Ricci coefficients of the null standard frame $T-N,T+N,e_2, e_1$ are given by
%the following
%\bea
%\chi^{\prime}_{AB} &=& \theta_{AB} - \eta_{AB} \\
%\underline{\chi}^{\prime}_{AB} &=& - \theta_{AB} - \eta_{AB} \\
%\underline{\xi}^{\prime}_A &= &\phi^{-1} \nlap_A \phi - a^{-1} \nlap_A a \\
%\underline{\zeta}^{\prime}_A &= &\phi^{-1} \nlap_A \phi - \epsilon_A  \\
%\zeta^{\prime}_A &= &\phi^{-1} \nlap_A \phi + \epsilon_A  \\
%\nu^{\prime} &= & - \phi^{-1} \nlap_N \phi + \delta  \\
%\underline{\nu}^{\prime} &= &  \phi^{-1} \nlap_N \phi + \delta   \\
%\omega^{\prime} &=&\delta -a^{-1} \nlap_N a
%\eea
%We use $\chi$, $\underline{\chi}$, etc for the  Ricci coefficients of the null frame
%$a^{-1}(T-N),a(T+N),e_2, e_1$.
%
%

\begin{Def}
We define the null components of the Weyl curvature $W$ as follows: 
\bea
\underline{\alpha}_{\mu \nu} \ (W) \ & = & \ 
\Pi_{\mu}^{\ \rho} \ \Pi_{\nu}^{\ \sigma} \ W_{\rho \gamma \sigma \delta} \
e_3^{\gamma} \ e_3^{\delta} 
\label{underlinealpha} \\ 
\underline{\beta}_{\mu} \ (W) \ & = & \ 
\frac{1}{2} \ \Pi_{\mu}^{\ \rho} \ W_{\rho \sigma \gamma \delta} \  e_3^{\sigma} \
e_3^{\gamma} \ e_4^{\delta} 
\label{underlinebeta} \\ 
\rho \ (W) \ & = & \ 
\frac{1}{4} \ W_{\alpha \beta \gamma \delta} \ e_3^{\alpha} \ e_4^{\beta} \
e_3^{\gamma} \ e_4^{\delta} 
\label{rho} \\ 
\sigma \ (W) \ & = & \ 
\frac{1}{4} \ \ ^*W_{\alpha \beta \gamma \delta} \ e_3^{\alpha} \ e_4^{\beta} \
e_3^{\gamma} \ e_4^{\delta} 
\label{sigma} \\ 
\beta_{\mu}  \ (W) \ & = & \  
\frac{1}{2} \ \Pi_{\mu}^{\ \rho} \ W_{\rho \sigma \gamma \delta} \ e_4^{\sigma} \
e_3^{\gamma} \ e_4^{\delta} 
\label{beta} \\ 
\alpha_{\mu \nu} \ (W) \ & = & \ 
\Pi_{\mu}^{\ \rho} \ \Pi_{\nu}^{\ \sigma} \ W_{\rho \gamma \sigma \delta} \
e_4^{\gamma} \ e_4^{\delta}  \ . 
\label{alphaR}
\eea
\end{Def}
As we are dealing with the Einstein vacuum equations, the Riemannian curvature tensor $R_{\alpha \beta \gamma \delta}$ is identically the Weyl curvature tensor $W_{\alpha \beta \gamma \delta}$. 

Thus we have the following with the capital indices taking the values $1,2$: 
\bea
R_{A3B3} \ & = & \ \underline{\alpha}_{AB} \label{intnullcurvalphaunderline*1} \\ 
R_{A334} \ & = & \ 2 \ \underline{\beta}_A \\ 
R_{3434} \ & = & \ 4 \ \rho \\ 
\ ^* R_{3434} \ & = & \ 4 \ \sigma \\ 
R_{A434} \ & = & \ 2 \ \beta_A \\ 
R_{A4B4} \ & = & \ \alpha_{AB}  \label{intnullcurvalpha*1}
\eea
with \\ 
\begin{tabular}{lll}
$\alpha$, $\underline{\alpha}$ & : & $S$-tangent, symmetric, traceless tensors \\ 
$\beta$, $\underline{\beta}$ & : &  $S$-tangent $1$-forms \\ 
$\rho$, $\sigma$ & : & scalars \ . \\ 
\end{tabular} 
\\ \\ 

Let $\tau_-^2=1+u^2$ and $r(t,u)$ is the area radius of the surface $S_{t,u}$. \\

The proof by Christodoulou and Klainerman in \cite{sta}, thus for situation (CK), yields the decay behavior: 
\beas 
\underline{\alpha}(W) \ & = & \ O \ (r^{- 1} \ \tau_-^{- \frac{5}{2}}) \\ 
\underline{\beta}(W) \ & = & \ O \ (r^{- 2} \ \tau_-^{- \frac{3}{2}}) \\ 
\rho(W) \   & = & \ O \ (r^{-3})  \\ 
\sigma(W)  \ & = & \ O \ (r^{-3} \ \tau_-^{- \frac{1}{2}}) \\ 
\alpha(W) , \ \beta(W) \ & = & \ o \ ( r^{- \frac{7}{2}})  \\ 
\eeas

The proof by the present author in \cite{lydia1, lydia2}, thus for situation (B), yields the decay behavior: 
\beas
\underline{\alpha} \ & = & \ O \ ( r^{- 1} \ \tau_-^{- \frac{3}{2}}) \\ 
\underline{\beta} \ & = & \ O \ ( r^{- 2} \ \tau_-^{- \frac{1}{2}}) \\ 
\rho , \ \sigma , \ \alpha , \ \beta \ & = & \ o \ (r^{- \frac{5}{2}})  \\ 
\eeas

The shears $\widehat{\chi}$, $\underline{\widehat{\chi}}$ with respect to the null vectorfields $e_4$ and $e_3$ generating the corresponding outgoing, respectively incoming null hypersurfaces (``light cones") are the traceless parts of the second fundamental forms defined as follows: 
Let $X, Y$ be arbitrary tangent vectors to $S_{t, u}$ at a point in this surface. Then
the second fundamental forms are defined to be 
\[
\chi (X, Y) = g(\nabla_X e_4, Y) \ \, \ \ \underline{\chi} (X, Y) = g(\nabla_X e_3, Y) . 
\]
The shears satisfy the following equations on $S_{t,u}$. 
\bea
\dlap \hat{\underline{\chi}} & = & \underline{\beta} + \hat{\underline{\chi}} \cdot \zeta 
+ \frac{1}{2} ( \nlap tr \underline{\chi} - tr \underline{\chi} \zeta )  \label{dchibund1} \\ 
\dlap \hat{\chi} & = & - \beta - \hat{\chi} \cdot \zeta + \frac{1}{2} ( \nlap tr \chi + tr \chi \zeta )  \label{dchib1}
\eea
where $\zeta$ denotes the torsion-one-form.

\section{Scri}

Along the null hypersurfaces $C_{u}$ as $t\rightarrow \infty $, it is true in both cases (CK) as well as (B) that 
\begin{equation}
\lim_{C_{u},t\rightarrow \infty }\left( rtr\chi \right) =2,\, \, \ \ \, \ \
\ \ \ \ \ \ \ \ \, \ \ \, \, \ \ \, \ \, \, \lim_{C_{u},t\rightarrow \infty
}\left( rtr\underline{\chi }\right) =-2  \label{asymptotic behaviour of trxi}
\end{equation}

\subsection{Situation (CK)}

On any null hypersurface $C_{u}$, the normalized curvature components $r%
\underline{\alpha }$, $r^{2}\underline{\beta }$, $r^{3}\rho$, $r^{3}\sigma$,  have limits as $t\rightarrow \infty $, in particular 
\begin{eqnarray*}
\lim_{C_{u},t\rightarrow \infty }r\underline{\alpha } 
&=&A \left( u,\cdot \right) ,\, \\ 
\lim_{C_{u},t\rightarrow \infty }\,r^{2}\underline{\beta } 
&=& B\left( u,\cdot \right) \\
\lim_{C_{u},t\rightarrow \infty }r^{3}\rho &=&P \left(
u,\cdot \right) ,\, \\
\lim_{C_{u},t\rightarrow \infty}r^{3}\sigma  & =& Q \left( u,\cdot \right) \\
\end{eqnarray*}
with $A$ a symmetric traceless covariant 2-tensor, $B$ a 1-form and $P$, $Q$ functions on $%
S^{2}$ depending on $u$. Denote by $\overline{P}$, respectively $\overline{Q}$, the mean values 
of $P$, respectively $Q$, on $S^2$. 
The following decay properties hold: 
\begin{eqnarray*}
\left| A \left( u,\cdot \right) \right| &\leq &C\left( 1+\left| u\right|
\right) ^{-5/2}\, \, \\ 
\left| B \left( u,\cdot
\right) \right|  & \leq & C\left( 1+\left| u\right| \right) ^{-3/2} \\
\left| P \left( u,\cdot \right) -\overline{P} \left( u\right) \right|
&\leq & C \left( 1+\left| u\right| \right) ^{-1/2}\, \, \\
 \left| Q \left( u,\cdot \right) -\overline{Q} \left( u\right)
\right| & \leq & C \left( 1+\left| u\right| \right) ^{-1/2} \\
\end{eqnarray*}
%and 
%\begin{equation*}
%\lim_{u\rightarrow -\infty }\overline{P}_{W}\left( u\right) =0,\, \ \ \ \ \
%\ \lim_{u\rightarrow -\infty }\overline{Q}_{W}\left( u\right) =0.
%\end{equation*}

From the structure equations one has, 
\bea
\nlap_N \hat{\underline{\chi}} & = &  \frac{1}{2} \underline{\alpha} + l.o.t.  \\ 
\nlap_N \hat{\chi} & = &  \frac{1}{4} tr \chi \hat{\underline{\chi}}  + l.o.t.  
\eea
Taking into account that far away from the source, namely on each $C_u$ as $t \to \infty$ the lapse $a$ tends to $1$, the following holds at large distances from the source of the radiation: 
\bea
\frac{\partial}{\partial u} \hat{\underline{\chi}} & = &  \frac{1}{2} \underline{\alpha} + l.o.t.  \label{rel1} \\ 
\frac{\partial}{\partial u} \hat{\chi} & = &  \frac{1}{4} tr \chi \hat{\underline{\chi}}  + l.o.t.   \label{rel2}
\eea

On the null hypersurface $C_{u}$, the shear $r^{2} \widehat \chi$ (as well as the normalized shear) 
has a limit as $t\rightarrow \infty $: 
\begin{equation*}
\lim_{C_{u},t\rightarrow \infty }r^{2}  \widehat \chi = \Sigma \left( u,\cdot
\right)
\end{equation*}
with $\Sigma $ being a symmetric traceless covariant 2-tensor on $S^{2}$
depending on $u$.
\begin{equation*}
- \frac{1}{2} \lim_{C_{u},t\rightarrow \infty } r \widehat{\underline{\chi}} 
 =   \lim_{C_{u},t\rightarrow \infty }r\widehat{\eta }=\Xi \left( u,\cdot \right)
\end{equation*}
with $\Xi $ being a symmetric traceless 2-covariant tensor on $S^{2}$
depending on $u$ and having the decay property 
\begin{equation*}
\left| \Xi \left( u,\cdot \right) \right| _{\overset{\circ }{\gamma }}\leq
C\left( 1+\left| u\right| \right) ^{-3/2}.
\end{equation*}
Moreover, the following relations hold (as a consequence of equations (\ref{rel1}) and (\ref{rel2}) and the asymptotic properties of the spacetimes): 
\begin{eqnarray}
\frac{\partial \Xi }{\partial u} &=&  - \frac{1}{4} A   \label{Xiu*1} \\ 
\frac{\partial \Sigma }{\partial u} &=& -    \Xi   \label{Sigmau*1}  . 
\end{eqnarray}

\subsection{Situation (B)}

On any null hypersurface $C_{u}$, the following limits are attained: 
\bea
\lim_{C_u, t \to \infty} r \underline{\alpha} & = & A(u, \cdot)
 \,,
   \\
\lim_{C_u, t \to \infty} r^2 \underline{\beta} & = & B(u, \cdot)
 \,,
  \label{23XI16.1}
\end{eqnarray}
with
\bea
| A(u, \cdot) | & \leq & c (1 + |u|)^{- \frac{3}{2} } \label{A}
 \,,
   \\
| B(u, \cdot) | & \leq & c (1 + |u|)^{- \frac{1}{2} } \label{B}
 \, .
\end{eqnarray}

Moreover, from (\ref{dchibund1}) it follows that the next relation holds even in this general case: 
\be \label{X123}
B = - 2 \dlap \Xi 
\ee

\section{Gravitational Radiation}

Working with the asymptotics of the situation in {(CK)}, proven by Christodoulou and Klainerman in \cite{sta}, Christodoulou derives in \cite{chrmemory} the null and ordinary memory. 
The details for Christodoulou's work are given in \cite{chrmemory} and the asymptotic ingredients are derived in the last chapter of \cite{sta}. 

From the asymptotics of the situation (B), proven by the present author in \cite{lydia1}, \cite{lydia2}, it follows that the memory integral diverges. We will consider a third situation treated by the present author in \cite{lydia3} and call it (B2). This is more general than (CK) but has more decay than (B) and a finite memory can be computed.

In {(B)}, due to the slow decay of the data, a certain integral in $u$ diverges that is the heart to compute memory. In view of this, note that the null limit of the $r^{-1}$ piece of the curvature behaves like in (\ref{A}). 
%\[
%| A(u, \cdot) |  \leq  c (1 + |u|)^{- \frac{3}{2} } 
%\]
Integrating w.r.t. $u$ twice gives a growth in $u$. Other important quantities do not have limits at null infinity. In particular, the shear $\hat{\chi}$ does not have a null limit. For these reasons, 
the memory formula at null infinity cannot be finite. 
(B) is an interesting borderline case in many ways: First, error terms have to be controlled in order to establish the stability result, many of which would diverge if the decay of the initial data were to be relaxed by a small epsilon. These are the borderline error estimates. 
Second, at null infinity we encounter different behavior for many interesting quantities. 
However, the Bondi energy is finite and non-negative. The integral over $u$ of the news tensor in the energy formula is finite but borderline. 

In order to further investigate the nature of gravitational wave memory, let us consider (B2) that lies between {(CK)} and {(B)}. Thus, stability of such systems is implied by the result {(B)}, but the asymptotics are different and more information can be retrieved from this situation. The data for {(B2)} are as follows: In definition \ref{intAFB} above replace the data by the following 
\bea
\bar{g}_{ij} \ & = & \ \delta_{ij} \ + \ l_{ij} \ + \ 
o_3 \ (r^{- \frac{3}{2}}) \label{afgeng2}  \\
k_{ij} \ & = & \ o_2 \ (r^{- \frac{5}{2}})    \ ,     \label{afgenk2}  
\eea
with $l_{ij}$ being homogeneous of degree $-1$, that is non-isotropic. Again, these are consequences of other weighted Sobolev norms to be controlled. In particular, it follows from an analysis of the Einstein equations together with the Bel-Robinson energies controlled in {(B2)} that the most important curvature components as well as the shears $\hat{\chi}$ and $\hat{\underline{\chi}}$ reach finite limits at null infinity. Consequently, all the terms in the memory formula are finite and the computation of memory is straightforward. 

Also note that relations (\ref{Xiu*1})-(\ref{Sigmau*1}) still hold for (B2). 

Thus, we have introduced three classes of spacetimes for which stability proofs have been established. We are going to investigate their structures to derive insights on the nature of gravitational memory. 

Next, we will explain what happens in the situations {(B)} and {(B2)} and compare these to {(CK)}. 

In all these situations, 
we can derive the following ($\dlap$-$\clap \ \ $) system on $S_{t,u}$ for the torsion $1$-form $\zeta$. Here, $\underline{\mu}$ denotes the conjugate mass aspect function. 
\bea
\clap \ \  \zeta & = & \sigma - \frac{1}{2} \hat{\chi} \wedge \underline{\hat{\chi}}   \label{z1} \\ 
\dlap \zeta & = & \underline{\mu} + \rho - \frac{1}{2} \hat{\chi} \cdot \underline{\hat{\chi}} \label{z2}
\eea 
Note that $\rho$ is the component $E_{NN}$ of the electric part of the Weyl tensor, and $\sigma$ the $H_{NN}$-component of the magnetic part. 
Whereas in the situation of {(CK)} and {(B2)} this system (multiplied by $r^3$) has well-defined limits at null infinity, in the situation of {(B)} that is not the case, because the leading order behavior is $r^{- \frac{5}{2}}$. Thus we cannot directly work with these. However, something else can be done and the system's lower order terms can be investigated more closely. 

The following different treatment works for all three cases {(B)}, {(B2)} and {(CK)}.

Consider the highest order terms in the Bianchi equations for $\underline{\beta}_3$ in (\ref{b3}) and $\rho_3$ in (\ref{r3}). 
Note that in \cite{sta}, \cite{lydia1}, \cite{lydia2}, \cite{lydia3} the full equations are used and $\underline{\beta}_3$ as well as $\rho_3$ involve lower order terms, whereas in the following it is simply $\underline{\beta}_3 = D_3 \underline{\beta}$ and $\rho_3 = D_3 \rho$. 
\be \label{b3}
\underline{\beta}_3 \  =  \ 
- \dlap \underline{\alpha} 
\ee
\be \label{r3}
\rho_3 \  =  \ 
 - \dlap \underline{\beta} 
 - \frac{1}{2} \hat{\chi} \cdot \underline{\alpha} 
\ee
For all {(B)}, {(B2)} and {(CK)}, (\ref{b3}) is $O(r^{-2})$. The situation is different for (\ref{r3}). In {(CK)} and {(B2)} all the terms in (\ref{r3}) are $O(r^{-3})$. In {(B)} we have for (\ref{r3})  
\[
\rho_3 \  =  \ 
 - \underbrace{ \dlap \underline{\beta} }_{= O(r^{-3})}
 - \underbrace{ \frac{1}{2} \hat{\chi} \cdot \underline{\alpha} }_{= O(r^{- \frac{5}{2}} u^{- \frac{3}{2}})}
\]
A short computation shows that 
\[
\rho_3 \  =  \ 
 - \underbrace{ \dlap \underline{\beta} }_{= O(r^{-3})} 
- \underbrace{\frac{\partial}{\partial u} (\hat{\chi} \cdot \hat{\underline{\chi}})}_{ = O(r^{- \frac{5}{2}} u^{- \frac{3}{2}})} + \underbrace{\frac{1}{4} tr \chi  |\hat{\underline{\chi}}|^2 }_{ = O(r^{-3})}
\]
Thus 
\[
\rho_3  + \frac{\partial}{\partial u} (\hat{\chi} \cdot \hat{\underline{\chi}})  \  =  \ 
 - \dlap \underline{\beta} + \frac{1}{4} tr \chi  |\hat{\underline{\chi}}|^2 \ =  \ O(r^{-3})
\]
Then we multiply this equation by $r^3$ and take the limit as $t \to \infty$ on $C_u$. We call $L_l$ the corresponding limit on the left hand side and obtain the asymptotic equation 
\be \label{Ll1}
L_l = - \dlap \underline{B} + 2 | \Xi |^2 
\ee
$L_l$ is a function on $S^2$ consisting of terms depending on $u$. 
Integrating with respect to $u$, noting that the contribution to $L_l$ from the shears goes to zero as  $| u | \to \infty$. 
The resulting term after integration we denote by $P$. 
In the cases (CK) and (B2) 
we obtain 
\be \label{PM1}
 -( P^+ - P^-) = - \dlap \int_{- \infty}^{+ \infty} \underline{B}  \ du \  + \   \int_{- \infty}^{+ \infty} | \Xi |^2 \ du 
\ee
where all the terms are well defined and finite. 
In the situation {(B)}, if one takes the integral with respect to $u$ of equation (\ref{Ll1}), then the integral of the first term on the right hand side diverges, the integrand being of order $|u|^{ - \frac{1}{2}}$. The integral of the last term on the right hand side is finite, in fact it is borderline. 
Next, we consider the cases (CK) and (B2). From (\ref{PM1}) we compute the null as well as the ordinary memory of gravitational waves as follows. 
Using (\ref{X123}) as well as relations (\ref{Xiu*1}) and (\ref{Sigmau*1}), we have 
\be \label{PM2}
\dlap \dlap (\Sigma^- - \Sigma^+) \ = \ (P^- - P^+) \ - \ \int_{- \infty}^{+ \infty} | \Xi |^2 du
\ee
On $S^2$ we define the function 
\be \label{Fen1}
F = \frac{1}{2} \int_{- \infty}^{+ \infty} | \Xi |^2 du 
\ee
Then $\frac{F}{4 \pi}$ is the total energy radiated to infinity per unit solid angle in a given direction. 
We also obtain the Bondi mass loss formula in a straightforward manner where $|\Xi|^2$ is integrated over $S^2$. 
From (\ref{b3}) we obtain
\[
\underline{B} (u) = - \dlap  \int_{- \infty}^{u} \underline{A} (u') \ du'
\]
From the Bianchi equation 
\be \label{s313}
\sigma_3 \ = \ - \clap \ \ \underline{\beta} - \frac{1}{2} \hat{\chi} \cdot \ ^* \underline{\alpha} + \cdots 
\ee
and the properties of the terms involved as \\ 
$|u| \to \infty$ 
we have 
\[
\clap \  \int_{- \infty}^{\infty}  \underline{B} \ du = 0 
\]
We find that there exists a function $\Phi$ such that 
\bea
\dlap (\Sigma^- - \Sigma^+) & = & \nlap \Phi   \label{ma1}  \\ 
\dlap \dlap (\Sigma^- - \Sigma^+) & = & \slap \Phi \nonumber  \\ 
&  \ = \ & (P - \bar{P})^- - (P - \bar{P})^+    \nonumber \\ 
& &  - 2 (F - \bar{F}) \label{ma2}
\eea
where barred quantities denote mean value on $S^2$. 
$\Phi$ has vanishing mean on $S^2$ and its projection to the first eigenspace of $\slap$ on $S^2$ is zero, which are precisely the intergrability conditions for the system. Thus, by 
Hodge theory we solve the system. 
Through the geodesic deviation equation and relations (\ref{Xiu*1}) and (\ref{Sigmau*1}) the quantity 
$(\Sigma^- - \Sigma^+)$ in (\ref{ma2}) multiplied by a factor including the initial distance of the test masses encodes the permanent displacement of test masses given by the ordinary and the null memory, giving the well-known memory formula. Note that we use a different convention for the optical scalar $u$ than in \cite{sta} that leads to opposite signs. 
Thus the solution of ((\ref{ma1}), (\ref{ma2})) gives the electric parity memory. 
This is consistent with the results of \cite{chrmemory} for spacetimes with fall-off properties as in (CK). 
For both classes of spacetimes (CK) and (B2) the memory given by ((\ref{ma1}), (\ref{ma2})) has two parts, namely the null memory sourced by $F$ and the ordinary memory sourced by $P$. For the Einstein vacuum equations $F$ is given by (\ref{Fen1}), and $P$ is the null limit of $r^3 \rho$ as explained above. 

Equations ((\ref{ma1})-(\ref{ma2})) can also be obtained using the system 
((\ref{z1})-(\ref{z2})) for the torsion-$1$-form $\zeta$. This was done by D. Christodoulou in \cite{chrmemory} for spacetimes of type (CK). In particular, the computation uses the relation between $\Sigma$ and $Z$ for the left hand side as well as the relation between the null limit of the conjugate mass aspect function $\underline{\mu}$ and the shear at null infinity $\Xi$. We will make use of that in the following part.

What about magnetic parity memory? 
In order to answer this question, we investigate the Bianchi equations again and focus on the highest order terms in the equation for $\sigma_3$ as follows. Recall (\ref{s313}). Integrating with respect to $u$ from $-\infty$ to $+\infty$ and taking into account the behavior of the terms involved, this yields 
\[
\clap \  \int_{- \infty}^{\infty}  \underline{\beta} \ du = 0 
\]
For {(B)} the same problem occurs, namely that the main components do not have limits at null infinity. Therefore, the information cannot be retrieved in this case. Let us focus on {(B2)}. As this can be treated easily with the system ((\ref{z1})-(\ref{z2})) we consider this now. We recall: 
\beas
\clap \ \  \zeta & = & \sigma - \frac{1}{2} \hat{\chi} \wedge \underline{\hat{\chi}}    \\ 
\dlap \zeta & = & \underline{\mu} + \rho - \frac{1}{2} \hat{\chi} \cdot \underline{\hat{\chi}} 
\eeas
Denote $\lim_{C_u,  t \to \infty} r^3 \sigma = Q$ and $\lim_{C_u,  t \to \infty} r^3 \rho = P$. 
From the result (B2) it follows that these equations multiplied by $r^3$ have well-behaved limits at null infinity. When integrating the limits on $S^2$ we obtain the mean values. It turns out that $\bar{P}$ has (different) nonzero values for 
$u \to \infty$ respectively, $u \to - \infty$, but $\bar{Q} \to 0$ for $|u| \to \infty$. 

We use the fact that from this system we can directly derive ((\ref{ma1})-(\ref{ma2})) and solve it to obtain the memory formula. (See above.) 

We recall that $\rho$ is the electric and $\sigma$ the magnetic part of the Weyl tensor in these equations. Consider equation (\ref{z1}) with the magnetic part $\sigma$. The first thing to notice is that there is no term that would contribute to null memory. The second observation is that if there were ordinary memory of magnetic type, then $Q$ should have non-vanishing and different limits as $u \to - \infty$ respectively $u \to \infty$ much like we find it for $P$ in the equation for electric parity memory. 
That is not the case because $Q$, respectively $\sigma$, decays in $u$ for large $|u|$. 

The crucial observation of the system ((\ref{z1})-(\ref{z2})) is, that in all 3 cases {(B)}, {(B2)} and {(CK)} the shear $\underline{\hat{\chi}}$ decays in $u$. In the borderline case, {(B)}, we have $\underline{\hat{\chi}} = O(r^{-1} u^{- \frac{1}{2}})$. In {(B2)} it is $\underline{\hat{\chi}} = O(r^{-1} u^{- \frac{3}{2}})$. 
For {(B2)} and {(CK)} multiply equations ((\ref{z1})-(\ref{z2})) by $r^3$ and take its limits at null infinity. Then their mean values on the sphere $S^2$ at infinity are computed to be 
\bea
\bar{Q} & = & - \overline{\Sigma \wedge \Xi} \label{oQ1} \\ 
\bar{P} & = & - \overline{\mathcal{M}} - \overline{\Sigma \cdot \Xi}  \label{oP1}
\eea
where $\mathcal{M}$ is the limit of the conjugate mass aspect function $\underline{\mu}$, namely 
$\mathcal{M} = \lim_{C_u,  t \to \infty} r^3 \underline{\mu}$. Moreover, we have 
$\overline{\mathcal{M}} = 2M_B$ with $M_B$ denoting the Bondi mass. 
It also follows that ${\mathcal{M}}$ tends to limits ${\mathcal{M}} \to {\mathcal{M}}^+$ as $u \to \infty$ and 
${\mathcal{M}} \to {\mathcal{M}}^-$ as $u \to - \infty$, and we have 
${\mathcal{M}}^- - {\mathcal{M}}^+ = -2F$. 
With the decay in $u$ of the shear term $\Xi$ it follows from (\ref{oQ1}) that for all spacetimes of type {(B2)} or {(CK)} 
$\bar{Q} \to 0$ as $|u| \to \infty$. 
For the spacetimes of type (CK) it is a direct consequence from \cite{sta}, namely from the control of the energy using rotational vectorfields, that the magnetic curvature component $\sigma$, respectively its null limit 
$Q$ decays in $u$ as $|u| \to \infty$. It follows that 
$(Q - \bar{Q})^+ = (Q - \bar{Q})^- = 0$. 
Now, the spacetimes of type (B2), see (\ref{afgeng2})-(\ref{afgenk2}), are more general involving the non-isotropic term $h_{ij}$ in the metric. 
From \cite{lydia3} it follows that the behavior of $\rho$ and therefore its limit $P$ is different than in the situation (CK), but it does give the non-trivial limits as $|u| \to \infty$ 
and an electric memory as described above. However, it also follows from \cite{lydia3} that $\sigma$ and therefore its null limit $Q$ do decay in $u$ as $|u| \to \infty$. To see this, consider the following demonstration: 
As we do not have the rotational symmetry in this case, the decay of 
$\sigma$ follows from its relation to $k$. In particular, we have for Einstein-vacuum spacetimes on each spacelike hypersurface $H_t$ 
\be \label{ksigma}
(curl \ k)_{lm} \ = \ H_{lm}
\ee
where $H$ is the magnetic part of the Weyl curvature as defined in (\ref{Wma1}), and $k$ behaving as in (\ref{afgenk2}). 
Note that we can decompose this equation into its parts tangential to and orthogonal to the surfaces $S_{t,u}$, but the full details are not needed right now. 
The $NN$-component of equation (\ref{ksigma}) reads 
\be \label{ksigma2}
(curl \ k)_{NN} \ = \ \sigma
\ee
Now, we recall the initial data for the (B2) spacetimes, in particular the decay of $k$ at spatial infinity as required by (\ref{afgenk2}). It is 
shown in \cite{lydia3} that this behavior is preserved under the evolution by the Einstein equations. Thus, from the fall-off of $k$ in 
(\ref{afgenk2}) and equation (\ref{ksigma2}) it follows directly that $\sigma$ and therefore also its limit $Q$ decay in $u$ as $|u| \to \infty$ for spacetimes (B2). Moreover, $(Q - \bar{Q})^+ = (Q - \bar{Q})^- = 0$ for (B2). 
As a consequence, there cannot be any ordinary memory of magnetic type either.  

Even for spacetimes of type {(B)} with very slow fall-off where the null limits of $\sigma$, respectively $\rho$ do not exist, looking at equations ((\ref{z1})-(\ref{z2})) and the properties of the quantities involved, we conclude that $\bar{\sigma}$ decays in $u$. 

We conclude that magnetic parity memory does not exist in all these situations. \\

{\bf General Remark}: When assuming the smallness conditions, all the mass will be radiated away and there will be no ordinary memory, only null memory. As the null asymptotic behavior is largely independent from the smallness of the data, we can work with large data. In a typical merger of two black holes or neutron stars the objects will radiate away mass and momenta. And the final object will be left with the remaining mass and momenta. In these cases, we find null memory and ordinary memory. All memory is of electric parity only.

\section{Stress-Energy}

Finally, we say a few words about stress-energy if the Einstein equations are coupled to another system. 
 If ``regular" matter or energy is coupled to the Einstein equations, then stress-energy terms appear in the Hodge system above, but only the electric part contributes to memory. 
 
Note that for the general Einstein equations coupled to a common matter or energy field, the analog of equation (\ref{ksigma}) 
reads 
\be \label{ksigmaRic2}
(curl \ k)_{lm} \ = \ H_{lm} + \frac{1}{2} \epsilon_{lm}^{\ \ \ j} R_{0j}
\ee 
where via the Einstein equations the Ricci tensor on the right hand side is given by the stress-energy under consideration. 
However, the $NN$-component is still given by (\ref{ksigma2}). 

An interesting example of a stress-energy that would be ``unusual" was put forward by R. Wald and G. Satishchandran (personal communication), whereas null magnetic memory does not occur, this example suggests a magnetic ordinary memory.

\section{Asymptotically-Flat versus Cosmological Spacetimes}

The treatment above has been for asymptotically flat systems. From our results \cite{bgsty1, BGYmemcosmo1} it follows directly that also in the common cosmological settings gravitational wave memory has null and ordinary memory coming from the $\rho$-equation only. Thus, there is electric parity only.

\section{Conclusions}

We showed that it is an intrinsic feature of large classes of solutions of the Einstein equations describing the physical world that gravitational wave memory is of electric parity, that is, magnetic parity memory does not occur. This holds for the Einstein vacuum equations as well as it includes all known types of matter or energy, that produce a memory. 

General relativity has enjoyed considerable leaps forward in the last decades on all its frontiers, combining new insights from physics, astrophysics and mathematics. Recent experiments and observations have opened the gate to new physical data and new methods in mathematics, in particular in geometric analysis and numerical relativity, have made it possible to answer burning questions. In particular, the LIGO/Virgo detections mark the beginning of a new era where we gain information from the universe directly rather than via electromagnetic waves. Making full use of all these resources and combining the various techniques from different fields, ranging from experiment to pure mathematics, will be essential to solve the challenging problems in the future.

\section{Acknowledgments}
The author thanks Demetrios Christodoulou for helpful remarks and discussions. 
The author thanks the NSF and the Simons Foundation; the author was supported by NSF Grants No. DMS-1253149 and No. DMS-1811819 and 
Simons Fellowship in Mathematics No. 555809.

\end{document}